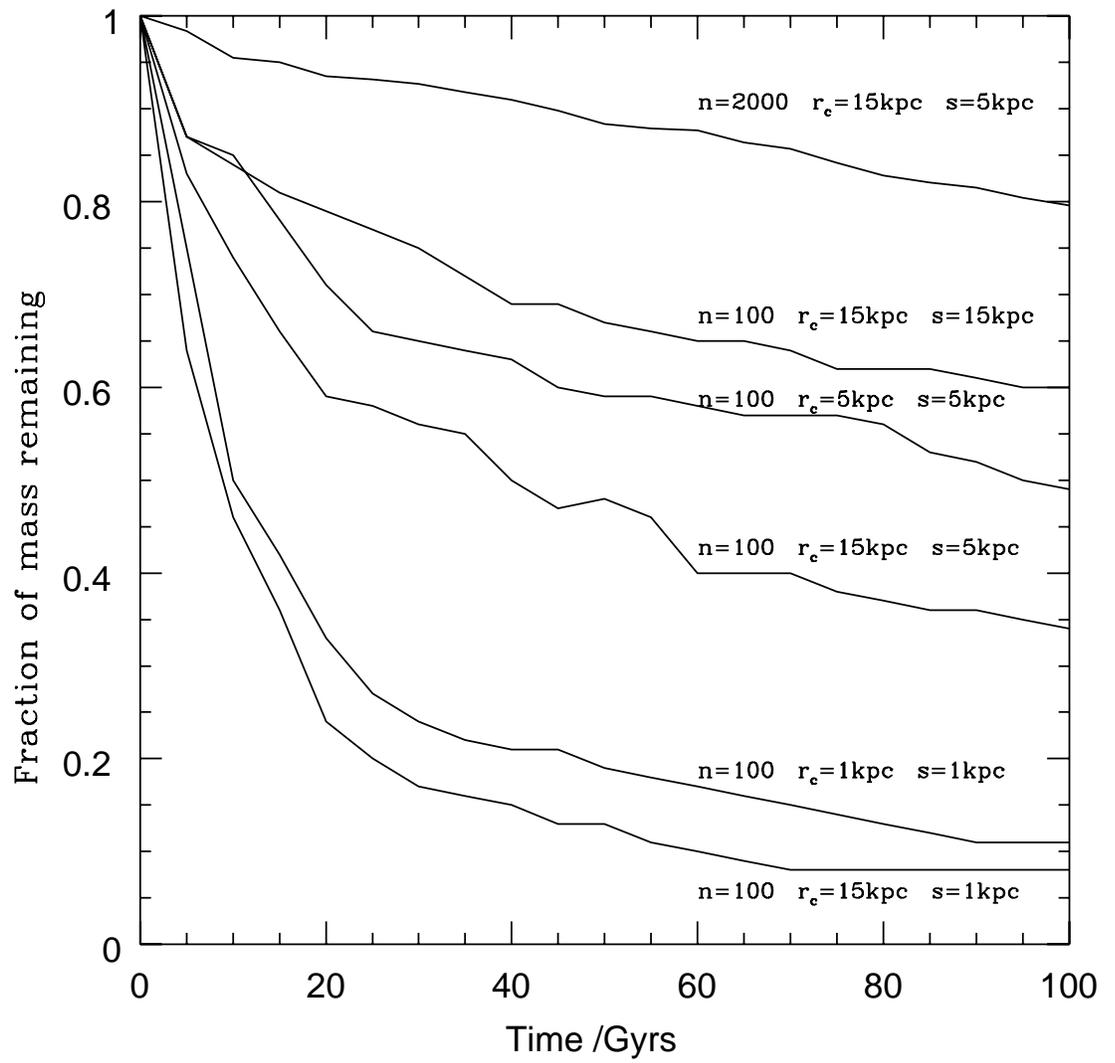

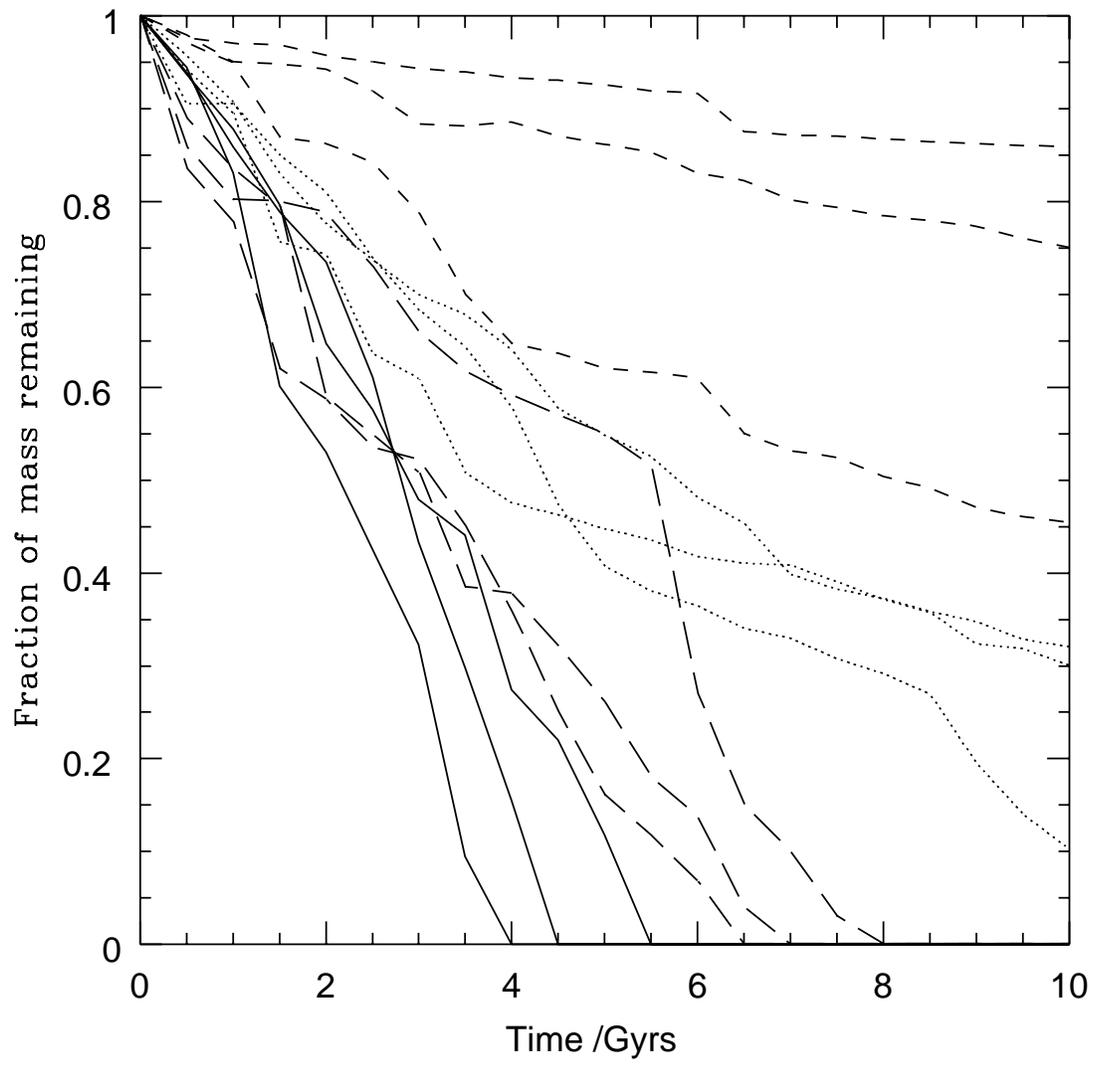

# On the destruction and over-merging of dark halos in dissipationless N-body simulations

Ben Moore, Neal Katz & George Lake

*Department of Astronomy, FM-20, University of Washington, Seattle, WA 98195, USA.*

# ABSTRACT

N-body simulations that follow only a collisionless dark matter component have failed to produce galaxy halos or substructure within dense environments. We investigate the 'over-merging' problem analytically and with numerical simulations, by calculating dissolution timescales of halos due to physical and artificial dynamical effects. The numerical resolution that has recently been attained is such that mass-loss from two-body relaxation is negligible. We demonstrate that substructure is destroyed in present simulations as a result of large force softening combined with the heating sources of tides and encounters with dissolving substructure. In the limit of infinite numerical resolution, whether or not individual halos or substructure can survive depends sensitively on their inner density profiles. Singular isothermal halos will always survive at some level, however, if halos form with large core radii then the over-merging problem will always exist within dissipationless N-body simulations. In this latter case a dissipational component can increase the halos central density enabling galaxies to survive.

**Subject headings:** Cosmology: Dark Matter – Large Scale Structure of Universe – Theory, Galaxies: Clustering – Formation – Interactions

## §1. INTRODUCTION

Understanding how structure evolves in the universe is a fundamental problem in cosmology. Cold dark matter dominated universes have been extensively investigated and have proven somewhat successful at reproducing observations from galactic to cluster scales (*e.g.* Davis *et al* 1985, White *et al* 1987). Structure formation within these models proceeds in a bottom up fashion, with small over-densities in the mass distribution collapsing first and subsequently merging hierarchically to form larger objects. Galaxies form from gas that has cooled within the deep potential wells provided by the dark matter halos and the accretion of gas rich sub-clumps that are part of the merging hierarchy (White & Rees 1978). The dominant force that drives structure formation is gravity, hence numerical simulations have proved a useful tool to study the non-linear growth of structure formation over a wide range of dynamical scales.

Several authors have attempted to study the formation of single rich galaxy clusters within a cosmological context in order to obtain very high spatial and mass resolution (Carlberg & Dubinski 1991, Carlberg 1994, Evrard *et al* 1994, Summers 1994). Their aim was to resolve individual halos in clusters at the present epoch. However, even with mass and length resolution $\sim 10^9 M_\odot$ and $\sim 15$ kpc respectively, no halos were found to survive within a distance of about half the virial radius from the cluster center. The so called 'over-merging' problem, or the inability to resolve sub-structure within dense environments, was first noted by White *et al* (1987) and Frenk *et al* (1988). This problem is most serious when comparing cosmological models with the observations. In order to create artificial galaxy catalogues it is customary to identify galaxies with over-dense peaks in the simulated mass distribution. i.e. peaks with 1-d velocity dispersion $\sim 150$ km s$^{-1}$ would be identified with $L_*$ galaxies. However, the cluster sized objects that actually form are devoid of any substructure or smaller peaks, leading investigators to artificially populate dense regions in the dark matter distribution with 'galaxies' (Davis *et al* 1985, Gelb & Bertschinger 1994).

Clusters in the real universe contain many galaxies that have retained their dark halos (at least within their optical radii). When the over-merging problem was first noticed, it was suggested that cooling gas would increase the central density of the halos, helping them stay intact within the cluster. Preliminary results from smoothed particle hydrodynamic simulations show that galaxy like clumps of gas and dark matter do survive to the present epoch (Katz *et al* 1992, Katz & White 1993, Evrard *et al* 1994, Summers 1994, Navarro *et al* 1995). These techniques are potentially very valuable, but are presently limited by their low resolution. Summers (1994) discuss different methods for identifying galaxy tracers within dissipationless simulations. They argue that the dynamics of the galaxy tracers identified as Carlberg (1994), is different from the dynamics of the clumps that form after a gaseous component is included.

In this *letter* we discuss the mechanisms that can cause dark halos to lose mass and dissolve in dense environments over a Hubble time. Artificial numerical effects include two-body relaxation of particles within the softened dark halos and the heating of halos by artificially massive cluster particles. Two physical heating effects will always be present, even in the limit of infinite numerical resolution: tidal heating of halos in eccentric orbits and impulsive heating from rapid encounters between halos. We shall use numerical simulations to follow the evolution of halos within a cluster environment in order to isolate these effects and test analytic estimates of the dissolution timescales. Our aim is to determine what erases substructure in present simulations and to determine whether or not future simulations at higher resolution will be able to resolve halos within dense environments.

## §2. DISRUPTION MECHANISMS

2.1 NUMERICAL EFFECTS

**2.1a Relaxation**

An $L_*$ galaxy halo with internal velocity dispersion $\sigma_h$ at a distance $R_c$ from a cluster with velocity dispersion $\sigma_c$, will have a tidally limited mass

$$m_h = 7.8 \times 10^{11} M_\odot \left(\frac{\sigma_h}{150 \text{ km s}^{-1}}\right)^3 \left(\frac{R_c}{500 \text{ kpc}}\right) \left(\frac{1000 \text{ km s}^{-1}}{\sigma_c}\right). \qquad (1)$$

Here we have assumed isothermal cluster and halo potentials so that the tidal radius of a halo is simply $r_t = R_c \sigma_h / \sigma_c$. Typical particle masses in cosmological simulations are between $10^9 M_\odot$ and $10^{10} M_\odot$, therefore the number of particles, $N_p$, within $r_t$ is in the range 100–1000. The evaporation time-scale from two-body encounters is of order $300 t_r$, where the median relaxation time-scale for a halo with half mass radius $r_h \approx r_t/3$ can be written $t_r = 0.14 N_p / \ln(0.4 N_p) \sqrt{r_h^3/(G m_h)}$ (Spitzer & Hart 1971), *i.e.* for isothermal potentials

$$t_{evap} = 3 \text{ Gyrs} \left(\frac{N_p}{\ln(0.4 N_p)}\right) \left(\frac{R_c}{500 \text{ kpc}}\right) \left(\frac{1000 \text{ km s}^{-1}}{\sigma_c}\right). \qquad (2)$$

Evaporation is accelerated in the presence of a tidal field (*c.f.* Chernoff & Weinberg 1990). To test the evaporation time-scale of a halo orbiting within a cluster, we constructed equilibrium halos with truncated isothermal density profiles. Our aim is simulate both the evaporation rate from halos that form in current simulations and those which might form given much better length resolution. Current cosmological simulations have force softenings of $\sim 5 - 20$ kpc and produce halos with correspondingly soft, low density cores of size comparable to the length resolution. Within the next few years a resolution of $\sim 1$ kpc will be obtained and the structure of individual halos should be resolved.

Each model was constructed with 100 particles with a total mass of $2.5 \times 10^{11} M_\odot$ within a physical radius of 30 kpc. The models were placed in circular orbits at 600 kpc in an isothermal cluster potential specified with a 1-d velocity dispersion of 1000 km s$^{-1}$. This distance was chosen so that the physical radius was well inside the tidal radius imposed by the cluster to minimise the effects of tidal heating. Figure 1 shows the fraction of mass remaining within a fixed distance of 30 kpc from the center of each halo over a period of 100 Gyrs. We find that the initial rate of mass loss is consistent with the standard evaporation formula above. However, the rate of mass loss is not linear with time, but rapidly slows down as the physical size of the halo approaches the softening or core size. Hence, halos with large softening and as few as 20 particles can survive many Hubble times. The rate of evaporation increases as the softening, hence core size, is reduced. *i.e.* individual encounters between particles can transfer more energy at the same impact parameter. Halos with the same softening will evaporate faster if the physical core is large since the binding energy is correspondingly lower.

## 2.1b Particle - halo heating

Carlberg (1994) explained the halo disruption in his $10^6$ particle cluster simulation as a consequence of heating by massive cluster particles. We can make an analytic estimate for $t_{ph}$ using the impulsive approximation, since the typical encounter time-scale is smaller than the galaxy's internal time-scale, *i.e.* $r_h/\sigma_h > 2b/\pi\sigma_c$, where $b$ is the impact parameter. This calculation was originally performed for the disruption of open clusters by giant molecular clouds (Spitzer 1958). As a cluster particle passes by or through a halo it tidally distorts the system and increases its internal kinetic energy. Following Binney & Tremaine (1987), we equate the disruption time-scale as the time for the halos binding energy, $E_{bind}$, to change by order of itself due to impulsive energy input:

$$t_d \approx \left(\frac{\Delta E}{E_{bind}}\right) = \left(\frac{0.03\sigma_c}{G}\right)\left(\frac{m_h}{r_h^3}\right)\left(\frac{r_p^2}{m_p^2 n_p}\right), \qquad (3)$$

where $m_p$ and $r_p$ are the perturbers mass and half mass radius (*i.e.* the Plummer softening length). Here, we have assumed that the galaxy's mean square radius is similar to the half mass radius $r_h$. Substituting the halos tidal radius for the half mass radius and noting that the number density of perturbers within an isothermal potential, $n_p = \sigma_c^2/(2\pi m_p R_c^2 G)$ we find

$$t_{ph} \approx 65 \text{ Gyrs } \left(\frac{\sigma_c}{1000 \text{ km s}^{-1}}\right)\left(\frac{r_p}{10 \text{ kpc}}\right)^2\left(\frac{10^9 M_\odot}{m_p}\right). \qquad (4)$$

The analogous analytic estimate for globular cluster dissolution by black holes in the halo of the Milky Way, was tested using N-body simulations by Moore (1993) and found to be accurate within a factor of order two. We therefore conclude that present dissipationless cluster simulations have obtained sufficient resolution to avoid this problem. *e.g.* Carlberg's (1994) cluster simulations has $m_p = 4 \times 10^9 M_\odot$ and $r_p = 20$ kpc, Evrard *et al* (1994) has $m_p = 10^9 M_\odot$ and $r_p = 10$ kpc.

2.2 PHYSICAL EFFECTS

**2.2a Tidal heating**

At a certain distance from the halo, particles will escape and become bound to the cluster potential. Our truncated softened isothermal halos will have somewhat smaller tidal radii than given by the above formula for true isothermal potentials but are more realistic halo potentials. We shall test the analytic formula by placing the model halos on circular orbits at different radii within the cluster. We define the tidal destruction radius to be the distance at which the halos loose 50% of their mass over 5 Gyrs. (Typical halos completely disrupt at this cluster-centric distance over a Hubble time.) Our model halo with core radius $r_c = r_p = 1$ kpc survives to $R_c = 200$ kpc. Halos with $r_c = 15$ kpc and $r_p = 5$kpc will survive to within about $R_c = 300$ kpc. Halos with $r_c = r_p = 15$ kpc can survive at $R_c = 450$ kpc.

Halos become very unstable when the core radius approaches the tidal radii and halos with fixed core radius will be more unstable with a larger softening. Hence, in the absence of all other disruption mechanisms, tidal disruption from the cluster potential is sufficient to erase all $L_*$ halos within 500 kpc for a Plummer softening of 10 kpc. *N.B.* the halo of an $L_*$ spiral galaxy will have a circular velocity $\sim 220$km s$^{-1}$(Tully & Fisher 1977). Note that the spline softening used here is equivalent to a Plummer softening about a factor of 30% smaller. Hence, simulations with 20 kpc Plummer softening will dissolve $L_*$ halos at about 1 Mpc within a Coma sized cluster! Any additional source of heating will drive this survival radius even further outwards.

Weinberg (1994) has stressed the importance of resonant orbit coupling between stellar dynamical systems orbiting within a tidal field. The orbital time of a halo at half the clusters virial radius is of order $10^9$ years, a time-scale similar to the internal orbital period within an $L_*$ halo at its half mass radius. Halos on eccentric orbits within a cluster will lose mass as a consequence of heating by the tidal field, even though the 'tidal shock' varies quite slowly with time. We test the importance of this effect numerically by placing our model halos in eccentric orbits within the same cluster potential. Typically between 10 and 20% of the mass is lost due to the slow tidal shocks over a period of about 10 Gyrs, halos with larger cores loose more mass than more tightly bound systems.

**2.2b Halo-halo heating**

An additional heating term arises from impulsive encounters between halos within the cluster. The rate of energy input via impulsive heating is $\propto m_p^2 n_p = f m_p \rho_c$, where $f$ is the fraction of the cluster mass within perturbers. Therefore the relative importance of heating from cluster particles versus individual halos is $f m_h / m_p$. Halos are typically between 10 and 100 times as massive as the particles in current simulations. Therefore we expect that the heating rate from halo-halo encounters will be about an order of magnitude more important than from particle-halo encounters.

Performing the same calculation for estimating $t_{ph}$, but replacing the softened particle perturbers with halos we find

$$t_{hh} \approx 3.5 \frac{1}{f} \text{ Gyrs} \left(\frac{R_c}{1 \text{ Mpc}}\right)\left(\frac{100 \text{ km s}^{-1}}{\sigma_h}\right). \tag{5}$$

Assuming tidally truncated isothermal halos, $f$ varies linearly with $R_c$ (at the cluster virial radius $f_{vir} \sim 0.5$) and we can write equation (5) as

$$t_{hh} \approx 5.3 \frac{1}{f_{vir}} \text{ Gyrs} \left(\frac{\sigma_c}{1000 \text{ km s}^{-1}}\right)\left(\frac{100 \text{ km s}^{-1}}{\sigma_h}\right). \tag{6}$$

The scaling of this formula shows that $t_{hh}$ is independent of position within the cluster. However, note that this derivation breaks down when the perturber mass is much larger than the halo mass and when the impact parameter is so large that use of the impulse approximation breaks down, *i.e.* when the encounter occurs beyond about 75 kpc from the halo. Hence, at large distances from the cluster center, very few encounters occur within this impact parameter and the heating rate falls to zero. Furthermore, in deriving this time-scale we have ignored the effect of the tidal field in estimating the binding energy of the halo. As we demonstrate below, halos closer to the center have lower binding energies and hence will be easier to disrupt. Thus we would expect halos to survive at the edge of clusters and to be disrupted with increasing ease towards the cluster center.

We test this analytic estimate by constructing a cluster of galaxies similar to the Coma cluster. Within the virial radius ($r_v/\text{kpc} = v_c/\text{kms}^{-1}$, where we have adopted $H_o = 100 \text{kms}^{-1}\text{Mpc}^{-1}$ and $\Omega = 1$) we place galaxies with luminosities drawn from a Schechter luminosity function parameterised using $\alpha = -1.25$ and $M_* = -19.7$, including all galaxies brighter than $2.8 \times 10^8 L_\odot$. Each cluster galaxy represents an isothermal potential, tidally limited at its pericentric distance, $R_c$, using $R_c \sigma_h / \sigma_c$. Here $\sigma_h$ is the velocity dispersion found using the Tully-Fisher relation (1977), converted from circular velocity assuming $\sigma = \sqrt{2} v_c$. This implies that our minimum luminosity corresponds to $\sigma_h = 50 \text{km s}^{-1}$. Since halos lose approximately half of their mass over a Hubble time and the impulsive energy input is proportional to the square of the perturber mass, we multiply the perturbers' velocity dispersion by 0.86. For $\sigma_c = 1000 \text{km s}^{-1}$ the cluster mass within the virial radius of 1.5 Mpc is $6.9 \times 10^{14} M_\odot$, hence for a mass to light ratio of 250, the total cluster luminosity within this radius is $2.8 \times 10^{12} L_\odot$. This normalization yields approximately 950 galaxies brighter than our minimum luminosity and 30 brighter than $L_*$. At 500 kpc from the cluster center, about 25% of the total cluster mass remains associated with the galaxies.

Our model halos are placed in circular orbits at 300 kpc and 600 kpc and feel the potential field of all the cluster members and the analytic potential of the remaining cluster background. Our treatment of the cluster environment contains all of the important dynamical effects. Its major shortcoming is that we assume the bulk of the cluster is in place at the beginning of the simulation. Figure 2 shows the results from this calculation. We find that the survival of a given halo depends sensitively on the ratio of the core radius to the tidal radius. As the core radius approaches the tidal radius the halos disrupt completely in a relatively short time-scale. Our estimate of the halo-halo disruption time-scale from equation (5) was in relatively good agreement with the numerical results for halos with small core radii.

The curves show stochastic evolution which leads to a dissolution time-scale of halos with fixed physical properties that varies by a factor $\sim 2$. The heating from a single encounter depends upon the square of the perturber mass, hence given a Schechter luminosity function, most of the heating is due to galaxies with luminosities $\gtrsim L_*/3$. This explains why the evolution is fairly chaotic since the number of massive perturbers is relatively small. Typically, the evolution of a halo is driven by between 5 and 10 large impulsive encounters.

## §3. DISCUSSION AND CONCLUSIONS

The resolution recently obtained within dissipationless N-body simulations is sufficient to resolve large halos within cluster environments. However, no substructure or halos are found within such environments in high resolution N-body simulations. We find that evaporation from numerical relaxation is negligible over a Hubble time for halos with more than 30 particles. However, present cosmological simulations have softening lengths of order $5-20$ kpc leading to halos with large low-density cores. As the limiting tidal radius approaches the halo core radius the dissolution of the halo occurs very rapidly. Hence, the over-merging problem within large over-dense regions is due to the large force softening combined with heating from the mean tidal field of the cluster and heating by encounters with remaining substructure. A second artificial numerical effect, particle-halo heating, does not pose a problem within numerical simulations as long as the particle mass is kept below $10^{10} M_\odot$.

In the limit of infinite numerical resolution, the survival of individual halos within dense environments depends entirely on their inner structure. If halos form with singular isothermal density profiles, then they can always be resolved at some level. Alternatively, including a gaseous component that dissipates energy can increase the central density of halos, effectively decreasing their disruption time-scale under halo-halo collisions. Recent results on the structure of dark matter halos give conflicting results (Moore 1995). The highest resolution simulations of Carlberg (1994), Warren *et al* (1992) and Crone *et al* (1994) give very steep inner density profiles. Furthermore, after correcting for force softening, Crone *et al* show that profiles are singular and fall steeper than $r^{-2.2}$ on all scales. These results disagree with those of Katz & White (1993) and Navarro *et al* (1994), who find that the density profiles of halos fall with radius as $r^{-1}$ over a large central region. If the former is true then it will be possible to identify halos within dissipationless simulations, although the halos themselves would be inconsistent with some observations of galaxy rotation curves (Moore 1994). If the simulated halos have shallow inner density profiles as some observations indicate, or as the latter authors find, then the over-merging problem can only be resolved by including a gaseous component.

## Acknowledgments

We thank Ray Carlberg for interesting discussions stimulated by his million particle cluster simulation. This work was funded by NASA through the LTSA and HPCC/ESS programs.

**Figure captions**

**Figure 1.** Mass loss rates from dark halos due to evaporation. The halos are constructed using the indicated model parameters for the core radius, $r_c$, and the softening length $s$, and particle number $n$. Each halo was placed on a circular orbit at 600 kpc from the center of an isothermal potential for 100 Gyrs.

**Figure 2.** Mass loss rates from dark halos due to halo-halo heating. For each model, circular orbits in different directions were followed to demonstrate the stochastic nature of halo-halo heating. The dotted curves show models with $r = s = 1$ kpc in orbit at 300 kpc. The solid curves show models with $r_c = s = 5$ kpc at 300 kpc. The short dashed curves show models with $r = s = 5$ kpc at 600 kpc. The long dashed curves show models with $r = s = 15$ kpc at 600 kpc.